# Should I Stay or Should I Go Now? An Investigation into Gender Differences in the Impact of Switching Jobs on Earnings

Emily Winskill

Under the supervision of

Sonia Oreffice

University of Exeter

May 2025

## Abstract

This paper investigates the relationship between job mobility and earnings growth in the UK labour market, with a focus on gender differences in the returns to switching jobs. Using data from the Annual Survey of Hours and Earnings (ASHE) between 2011 and 2023, the analysis compares wage progression for job switchers and stayers, controlling for individual and job characteristics. The findings show that job mobility is associated with higher earnings growth, but women experience smaller gains than men, with occupational mobility and age further widening this gap. However, the study finds no statistically significant evidence that changes in occupation, sector, or working time pattern influence this gender gap. The results highlight the importance of addressing gender disparities in the returns to job mobility and provide valuable evidence for developing policy interventions aimed at promoting more equitable labour market outcomes.

# 1: Introduction

The gender pay gap remains a persistent feature of the UK labour market despite decades of progress towards equality. According to the ONS, in 2024 the gender pay gap for full-time employees was 7%, with greater disparities amongst higher earners and in certain sectors (ONS, 2024). This continuing disparity suggests that structural and behavioural factors continue to perpetuate gender inequalities. Early research by Oaxaca (1973) highlighted the role of discrimination in wage disparities, and further studies have confirmed that structural and behavioural factors maintain these inequalities (Anderson, et al., 2001; Apergis & Lynch, 2022; McGee, et al., 2015). Despite efforts to narrow the gender pay gap through policy interventions, disparities persist.

A potentially important but underexplored mechanism contributing to this gap is the impact of switching jobs on men's and women's earnings growth. Job mobility has been shown to be a key determinant of wage growth, with studies showing that changing jobs often leads to substantial earnings gains and opportunities for career advancement and skill development (Cominetti, et al., 2022; Kirkup & Petrie, 2020). However, although much research has been done into the impact of switching jobs on earnings, and the gender pay gap, less has been done on the interaction of these two phenomena, to discover if men and women benefit equally from job switching. In the UK, approximately 2.4% of workers switch jobs each quarter (ONS, 2024), and switching is viewed as a key mechanism for workers to achieve opportunities for career progression, skill development, and access to higher wages. However, whether men and women benefit equally from this remains a question. If men consistently secure higher wage growth following a job switch, this could exacerbate existing gender inequalities in the labour market over time.

The UK labour market provides a particularly interesting context for examining this issue. Research consistently shows that the UK gender pay gap is smaller amongst younger workers, but expands over time (Manning & Swaffield, 2008; ONS, 2024). This suggests that job mobility, which typically increases with career progression, may play a critical role in the widening gender pay gap over time. Also, initial employment rates for men and women are relatively similar (ONS, 2025), and mobility is moderate by international standards (Van Ours, 1990). Understanding the extent to which gender influences the wage returns to job mobility is therefore critical for explaining the persistence of the gender pay gap.

Although the majority of literature suggests that gender differences exist in the wage returns to job switching (Avram, et al., 2023; ONS, 2022), the findings are not universally agreed upon. Some studies suggest that job mobility may benefit both genders equally under certain conditions (Keith & McWilliams, 1999). The ambiguity in these findings highlights the need for a closer examination of the specific conditions under which job switching either mitigates or exacerbates gender inequalities. It is essential to isolate the effects of gender from other factors, such as occupation, sector, and full-/part-time status to better understand the precise nature of these dynamics.

This study aims to contribute to the literature by analysing the gendered dynamics of wage growth through job switching in the UK, using recent longitudinal data from the Annual

Survey for Hours and Earnings (ASHE). This offers a representative sample of UK workers, enabling a detailed analysis of wage growth following job transitions. Unlike many previous studies that have treated gender as a control variable or descriptive feature, this research aims to provide a more comprehensive view by considering broader demographic and occupational factors that may affect earnings. Furthermore, the lack of up-to-date research on this topic in the UK makes it difficult to draw conclusions about this specific to the current dynamics of the UK labour market, particularly in light of recent economic shocks and labour market changes.

This research is especially relevant now, given the rapidly changing dynamics of the labour market. For example, the Covid-19 Pandemic had a disproportionate impact on women, who were more likely to experience furlough, job loss, and increased care responsibilities (Women and Equalities Committee, 2021). Since the pandemic, job switching rates have increased due to greater demand for work-life balance and rising inflation, contributing to what has been referred to as the "Great Resignation" (Gratton & Cable, 2022). In this context understanding the relationship between job switching, earnings and gender is crucial for informing effective policies aimed at closing the gender pay gap. The broader economic implications of gender differences in the wage returns to job switching are substantial. Disparities in wage growth from job mobility can result in reduced productivity and slower economic growth, ultimately affecting the broader economy. Addressing these inequalities could help to ensure a more equitable distribution of economic opportunities, benefitting both individuals and the wider economy.

In addition to addressing a gap in the literature, this research also has important policy implications. If job switching is found to benefit men more than women, or if the benefits are uneven across different sectors or demographic groups, targeted interventions may be required to ensure that women can fully access the wage growth opportunities associated with job mobility. This could involve policies aimed at reducing barriers to achieving the full wage growth potential from job changes, such as improving access to higher-paying opportunities or addressing gender biases in hiring and promotion processes. To effectively design these policies, it is necessary to understand not only the extent of the gap but also underlying factors that contribute to it, ensuring that women can fully benefit from the wage growth opportunities associated with job mobility.

# 2: Literature Review

## 2.1: Introduction

Whilst the gender pay gap has been widely studied, less attention has been paid to how this interacts with earnings growth from job mobility. Given the importance of job switching for career progression and wage advancement, understanding whether the benefits of this are equally distributed between men and women is critical for explaining how earnings disparities evolve over time. This literature review begins by outlining the broader research on the relationship between job switching and earnings. It then examines existing evidence on gender differences in the wage returns to job switching, considering many explanations for any observed disparities.

## 2.2: Trends in Job Switching and Earnings

Job mobility, the frequency with which people change jobs, has fluctuated since 2000, influenced by economic cycles and evolving workplace dynamics. According to data from the Office for National Statistics (ONS), the rate of job-to-job moves has generally risen since the early 2010s, dipping during recessions, and spiking during the post-pandemic recovery (ONS, 2024). However, moves within companies are not captured in this data, suggesting that job mobility is even more prevalent than previously suggested.

Demographic factors play a crucial role in mobility patterns. Younger workers, those in lower paid positions, and those in private sector service industries exhibit higher mobility rates (Cominetti, et al., 2022). However, rising living costs and economic uncertainty have recently reduced incentives to switch jobs for younger workers due to increased risk (Willetts, et al., 2018; Crowley, 2024). Mobility also responds to economic conditions, falling during recessions when job security becomes a priority and rising in tighter labour markets where workers have greater bargaining power (Strathmann, 2025).

Other factors shaping mobility include the growing importance of work-life balance, with 6% of workers in 2023 switching jobs due to a lack of flexible working options (CIPD, 2023). This shift reflects a change in perceptions of job switching, with less stigma and more emphasis on wellbeing. Inflationary pressures also impact job mobility, as high inflation encourages workers to seek higher paying roles to keep up with the cost of living (Pilossoph, et al., 2024).

Numerous studies have explored long-term trends in the effects of changing jobs on earnings, and have generally found that the effect is positive, as switching allows employees to negotiate higher salaries, gain promotions and move into higher paid industries. Research conducted by Cominetti et al. (2022) found that since 1975, job switchers have earned, on average, four percentage points higher pay growth than stayers. This premium is more pronounced in economic upturns and when workers move sectors or regions. However, involuntary job movers often experience negative wage growth, and the wage premium for job switchers shrinks after recessions.

However, the extent to which job switching leads to wage growth depends on several sectoral and demographic characteristics. Particularly in early-career stages, frequent job switches are associated with rapid wage growth (Kirkup & Petrie, 2020). As a very high amount of accumulated wage growth is concentrated in the first ten years, it is especially relevant (Murphy & Welch, 1990). Older workers are more settled into their careers, so use switches to generate gradual wage increases (Haltiwanger, et al., 2018). Workers who switch industry or occupation when changing jobs tend to experience higher wage growth than those who stay in the same field, especially for lower-wage workers and in economic upturns. However, during recessions, the wage changes from changing careers decline significantly, sometimes falling below those of career stayers (Carrillo-Tudela, et al., 2016). Higher-skilled and more educated workers benefit more from job mobility, whilst low-income workers receive low or even negative wage growth when switching jobs due to weaker bargaining positions and job insecurity (Cominetti, et al., 2022; Kirkup & Petrie, 2020). Furthermore, occupational segregation limits mobility opportunities, especially for those in female-dominated occupations, restricting their ability to move for higher wages (Perales, 2010). Given these constraints, job mobility does not lead to equal wage gains for all workers.

## 2.3: Theoretical Framework

Several theories explain why job switching can influence earnings differently by gender. Becker's (1964) Human Capital Theory suggests that job moves allow workers to build skills and knowledge that increase productivity and consequently wages. However, gender differences in human capital acquisition, such as women's greater likelihood of career interruptions and part-time work, can limit these gains.

Additionally, economic theory suggests that job switching reflects a process of job matching, where workers and firms continuously seek the best fit (Jovanovic, 1979). Employees may switch jobs if they receive offers that better align with their abilities, and firms offer higher wages to attract and retained skilled workers, creating a competitive labour market where mobility drives wage growth. Women may face higher frictions to their job search due to discrimination, structural and individual factors, reducing the wage returns to switching.

Furthermore, Rosen's (1986) theory of compensating differentials suggests that some workers may accept lower wages in return for other benefits, such as more flexibility, a pattern that may be more common amongst women due to greater household responsibilities.

Finally, the Dual Labour Market theory distinguishes between stable, high-pay 'primary' and unstable, low-wage 'secondary' labour markets (Klimczuk & Klimczuk-Kochańska, 2016). Women are disproportionately represented in secondary labour markets, limiting the wage growth achievable through mobility.

## 2.4: Gender Disparities in the Impact of Switching Jobs on Earnings

The relationship between gender, job mobility and earnings is complex, and in the UK, evidence on gender differences in switching rates is mixed. Several studies have found broadly similar rates of job mobility between men and women (Booth & Francesconi, 2000; Avram, et al., 2023; Cominetti, et al., 2022), although nuances exist. Booth and Francesconi (2000) found that women faced higher layoff rates, although their findings, limited to white, full-time workers, may no longer fully reflect modern labour market dynamics. Avram et al. (2023) found that mothers are more likely to switch jobs for family reasons rather than career reasons, with significant implications for wage growth. Cominetti et al. (2022) further observed higher rates of occupational upgrading amongst women. By contrast, Carillo-Tudela et al. (2016) found that women have higher rates of job mobility than men, suggesting that patterns may vary depending on the measures and contexts used. One potential consequence of higher job switching rates amongst women is reduced employer investment in training, limiting human capital accumulation and contributing to persistent gender pay disparities (Booth & Francesconi, 2000).

Findings from the US, however, point to a somewhat different dynamic. Several studies suggest that women, particularly young women, exhibit lower rates of job switching than men (Loprest, 1992), and that men are more likely to search for a new job whilst employed (Keith & McWilliams, 1999). Marriage and children have been shown to significantly reduce the likelihood of job searching for young women, but not for young men (Yankow & Horney, 2013). This pattern of lower job mobility amongst women is often attributed to traditional gender roles, work-family balance pressures, occupational segregation, and the prevalence of women in part-time and fixed-term roles which typically offer fewer promotion opportunities. Broader behavioural differences may also play a role, with women generally exhibiting greater risk aversion than men (Dawson, 2023), potentially making them less likely switch jobs frequently.

Together, these findings highlight that the relationship between gender and job mobility is highly context-dependent, shaped broader social, economic and behavioural factors. Whilst UK-based studies are the primary focus, insights from the US offer a valuable comparison, helping to contextualise how different labour market structures and social norms might influence gender differences in patterns of job switching. Understanding these patterns is crucial for assessing how the wage returns to job switching may differ between men and women.

Whilst there is considerable evidence that switching jobs tends to increase earnings growth for all individuals, less research has been done into the gender differences in the wage returns to mobility, and the available findings are mixed. Del Bono & Vuri (2006) find that in Italy, job mobility accounted for up to 30% of total wage growth for men over the first ten years of their careers, but only 8.3% for women, even after controlling for individual and firm characteristics. They attribute approximately half of the gender gap in wage growth to gender differences in wage returns to mobility, with moves to larger firms playing an important role. In the UK, the ONS (2022) found that in 2021, women switching jobs experienced earnings growth of 0.9 percentage points lower than that of men, compared to a gap of only 0.2 percentage points for stayers, with this pattern appearing to have been consistent over time. Similarly, Avram et al. (2023) found that differences in job mobility patterns explain around a quarter of the gender pay gap between men and women with similar characteristics.

However, other studies paint a more mixed picture. Keith and McWilliams (1999) found no significant difference in the wage premium associated with mobility between men and women, suggesting that switching may not always amplify gender disparities. Others argue that differences in job search behaviour, rather than the wage returns to mobility, play a greater role in explaining the gender pay gap lies in the job searching process (Manning, 2003). Cha (2014) further complicates the evidence, finding that voluntary job changes prior to the 2008 recession were associated with greater earnings growth for women than men, particularly amongst childless women. However, for involuntary moves, women experienced greater earnings losses than men.

Overall, whilst international evidence on gender differences in the wage returns to job mobility is mixed, findings from the UK tend to suggest that a disparity persists, with women benefitting less than men from job switching.

One potential explanation for the gender disparity in wage returns to job mobility lies in the gender differences negotiation behaviours. Research consistently suggests that women have lower bargaining tendencies, even when both groups are given the same careers advice, and women are less likely to negotiate salary unless the job advert specifically says that the salary is negotiable (Babcock & Laschever, 2007; Leibbrandt & List, 2015). Furthermore, Card et al. (2015) found that women gain less than men when moving to higher-paying firms, but lose less when switching to lower-paying firms, indicating weaker bargaining power when it comes to gaining firm-specific wage premiums. However, the evidence is not unanimous. Marks and Harold (2011) and O'Shea & Bush (2002) found no significant difference in the likelihood or success of negotiation. This suggests that whilst bargaining powers may contribute to gender disparities in earnings growth when switching jobs, they are not the sole reason.

Access to networks and social capital also appears to play an important role. Kronberg (2013) argues that women are often excluded from professional networks, limiting their opportunities for upward mobility. Family dynamics exacerbate this, as childcare responsibilities negatively affect women's networks more than men's (Campbell, 1988), undermining women's chances of benefitting from networks during job transitions.

Employer biases may also contribute to slower wage growth for women when switching jobs, with evidence suggesting that statistical discrimination affects hiring and promotion decisions, particularly when employers assume that women are more likely to prioritise family responsibilities (Avram, et al., 2023). In monopsonistic labour markets, where choice of employer is more limited, such biases may be even more damaging, further constraining women's wage progression.

Career interruptions further contribute to the difference in wage returns to mobility. Women are more likely than men to take employment breaks, especially for parenthood, which can erode firm-specific human capital. Gottschalk (2001) found that women experience greater wage losses than men following a job change involving time out of the labour force. Employer perceptions of reduced competence and commitment after career breaks may also influence wage offers (Ridgeway & Correll, 2004). Manning & Robinson (2004) similarly found that whilst early-career wage growth was comparable across genders, disparities emerged after breaks in employment, with men subsequently earning more. However, these findings should be interpreted cautiously given the small sample size used in this study.

These disparities are also evident within different segments of the labour market. Drawing on dual labour market theory, Kronberg (2013) distinguishes between 'good' jobs in the primary labour market and 'bad' jobs in the secondary labour market, finding that voluntary job changes in 'good' jobs are associated with a narrowing of the gender earnings gap, whereas in 'bad' jobs, women experience lower earnings gains to men. She found that voluntary job changes in good jobs are associated with narrowing gender earnings gap, whereas women don't experience as good gains as men. Conversely, involuntary moves in 'bad' jobs narrow the gender gap, whilst in 'good' jobs, they tend to widen it over time. This highlights that the gendered wage returns to job mobility are not uniform, but vary substantially between segments of the labour market.

A major reason for the gap is occupational and sectoral segregation. Women remain disproportionately concentrated in public sector, health, and education roles (Bosworth & Kersley, 2015) which typically offer lower wage returns to mobility. This segregation is particularly important given that industry switching is common, with only 35% of job changers between 2020 and 2022 remaining in the same industry (McKinsey, 2022). Research consistently shows that moves into male-dominated occupations bring higher wage growth than moves into female-dominated ones, where mobility is often motivated by family considerations rather than financial incentives (Rosenfeld & Spenner, 1992). Pearlman (2018) finds that men without college education in male-dominated occupations receive twice the wage growth as those in more gender-balanced or female-dominated occupations. However, for college-educated individuals, this gender gap largely disappears, suggesting that education partially mitigates, but does not eliminate, this disparity. However, Loprest (1992) found that specific occupational moves did not fully explain gender differences in wage growth after job changes, suggesting that structural barriers exist beyond occupational segregation. Overall, the literature mostly shows that occupational segregation limits opportunities available to women when switching jobs and reduces the financial rewards of mobility.

Part-time employment also plays an important role in limiting the wage gains women receive from job switching. Women are more likely to move into or remain in part-time roles, often due to caregiving responsibilities and a lack of flexible full-time positions (Francis-Devirle, et al., 2025; Avram, et al., 2023), and part-time positions typically offer lower wage growth and fewer promotion opportunities (Rubery, et al., 2024), reducing the benefits of mobility. During the 2008 financial crisis, women's earnings growth was initially less impacted, partially due to their overrepresentation in part-time and public sector roles, but female job changers experienced a slower recovery compared to men (ONS, 2019). Research consistently shows that part-time transitions account for a significant share of the gender gap in wage gains in job mobility. In the US, Loprest (1992) found that men gained, on average, twice as much as women from switching jobs, with transitions into part-time work accounting for 20% of this difference. Similarly, UK research shows that women who switch to part-time work experience negligible wage growth compared to those switching to full-time roles, with this effect particularly pronounced amongst higher educated women (Dias, et al., 2018). Overall, the literature seems to suggest that part-time status substantially limits potential earnings growth when changing jobs, particularly for women.

Age also shapes the outcomes of job switching. Younger workers generally experience higher wage growth from mobility (ONS, 2022), benefitting from lower firm-specific human capital losses and greater flexibility (OECD, 2024). However, women, especially mothers, often see smaller earnings benefits from early-career moves tend to be smaller, despite similar switching rates between genders (Manning & Swaffield, 2008; Avram, et al., 2023; Hospido, 2009).

Regional variation also affects labour market outcomes. Women in rural areas face fewer employment options, limiting their bargaining power and earnings potential from switching jobs. Lagerström and Eriksson (2008) found that women are less willing to accept jobs further from home, reducing their opportunities to access higher-paying roles. However, women who do expand their search area tend to receive more responses from employer, indicating that geographic constraints play an important role in the gendered outcomes from job switching, with women's opportunities more restricted by regional factors compared to men.

The structure of mobility (whether within or between firms), significantly affects gender disparities in wage outcomes. Studies consistently show that men tend to benefit more from external switches, particularly in the form of bonuses and commissions (Quintana-Garcia & Elvira, 2017; Brett & Stroh, 1997). In contrast, women switching firms often experience stagnant or lower earnings, even after controlling for tenure and industry. Internal moves, by contrast, tend to bring positive wage growth for both genders (Le Grand & Tåhlin, 2002), and may be particularly important for women, who face greater barriers in navigating external job changes. Female representation in leadership positions has been shown to mitigate these disparities (Quintana-Garcia & Elvira, 2017). However, evidence is mixed, as some studies suggest that external moves may help to close the gender gap for women already in high-paying roles (Kronberg, 2013), whilst broader trends indicate that external job changes disproportionately disadvantage women, reinforcing wage growth inequality.

Skill level further influences earnings growth from job mobility. Higher-skilled workers generally gain more from switching jobs, with educational attainment helping to reduce, but not eliminate, gender disparities (Pearlman, 2018; Cominetti, et al., 2022). Better educated workers are also more likely to change jobs, benefitting from a wider range of opportunities

(Amior, 2019). Whilst education mitigates some gender disparities in earnings growth from job mobility, it does not fully eliminate them, particularly for those in lower-skill roles.

Finally, sectoral segregation may also play a role. Women are significantly overrepresented in the UK public sector, comprising 65% of the workforce, compared to only 43% in the private sector (ONS, 2025). Private sector workers are more mobile (Cominetti, et al., 2022), and whilst private sector switchers were more affected by the 2008 financial crisis, they also experienced a faster recovery (ONS, 2019). However, limited research intro how sector dynamics interact with mobility outcomes for men and women leaves a gap in understanding the full picture.

## 2.5: Conclusion

The literature consistently shows that job switching tends to increase earnings growth, but with notable gender differences. Men generally experience higher wage returns from job mobility, whilst women's experiences are often influenced by structural factors such as part-time work, occupational segregation and the structure of mobility. Whilst existing studies have explored these dynamics, there remains a gap in understanding how gender differences in the return to job mobility differ across various demographic and sectoral groups. This research aims to address these gaps and capture more recent labour market dynamics. By including age, full-time versus part-time status, sector and occupation as factors, this study will offer a deeper understanding of the gendered impacts of job mobility.

# 3: Methodology

This analysis aims to identify whether the effect of job switching on earnings growth differs by gender, using data from the Annual Survey of Hours and Earnings (ASHE). The framework builds on previous studies (Avram, et al., 2023; Fuller, 2008), using a panel fixed effects regression model, with interaction terms and robustness checks to assess heterogeneity and the stability of the results.

## 3.1: Data Description

The ASHE dataset, collected by the Office for National Statistics (ONS), includes approximately 135000-190000 individuals per year, drawn from the PAYE records. It offers high reliability, consistent sampling methods, and less attrition than other sources (ONS, 2022), and includes detailed information on earnings, hours, occupation and employment characteristics. ASHE data are restricted, and access is subject to specific licensing agreements.

Covering the period from 2011 to 2023, the data excludes the 2008 financial crisis period, focusing instead on more recent labour market dynamics. The inclusion of the Covid-19 pandemic years is intentional, as it allows the study to capture how major economic shocks affect labour market outcomes. Previous literature often relies on relatively brief time periods or outdated studies, which may not capture the long-term effects of recent labour market shifts, including the impacts of the pandemic and the transition to a net-zero economy. This study therefore aims to address this gap. Only employees present in two consecutive years are included, enabling the identification of job switchers, defined as individuals who changed employer from one year to the next, and stayers, who remained with the same employer over two consecutive years.

Data cleaning removed duplicates, retained each individual's main job and excluded observations with zero or missing hourly earnings or pay affected by absence (e.g. sickness or maternity leave). After these steps, 1,337,929 person-year observations remained for descriptive analysis, and approximately 34% of observations were excluded due to data cleaning and sample restrictions for the regression analysis. Longitudinal weights were applied to correct for non-response bias. The dependent variable is hourly earnings growth, computed as the percentage change in hourly earnings (including overtime) between two consecutive years. Earnings growth is used rather than static earnings to allow for a direct assessment of the impact of job switching on wage progression. Nominal earnings are used rather than real as the dataset did not include price indices, and the relative disparities between groups remain meaningful in nominal terms.

Explanatory variables include a job switching and gender, and controls for age group, full-time versus part-time status, sector (public/private) and occupation (one-digit SOC). Many of these have not been included in past research, so including them should make the results more robust. Variable for changes in occupation, industry, and region were also generated. Certain important characteristics, such as ethnicity, education, skill level, parenthood status, and whether job changes were voluntary or involuntary, are not available in the dataset, which imposes limitations on the analysis.

The distributions of both hourly earnings and hourly earnings growth are highly skewed, justifying the use of median statistics rather than mean, as specified by the ONS (ONS, 2017), as medians are more robust to extreme outliers and better reflect the central tendency in heavily skewed wage data. In regression models, log-transformations of earnings growth are applied to normalise the distribution and facilitate the interpretation of coefficients as approximate percentage changes.

### 3.2: Summary Statistics

Table 1 presents the summary statistics for the variables included in the analysis.

*Table 1: Summary Statistics*

| Variable | Median | Mean | Std. Dev. | N |
|---|---|---|---|---|
| sex | 1 | 1.490527091 | 0.499910256 | 1337929 |
| sjd (job switch indicator) | 1 | 1.087838038 | 0.283059212 | 1337929 |
| he (Hourly Earnings) | 1368.588781 | 1699.671818 | 2667.483817 | 1337929 |
| heg (Hourly Earnings Growth) | 3.016089245 | 8.12374731 | 137.8076487 | 1337929 |
| age | 43 | 42.64564516 | 12.54645173 | 1337929 |
| ft (Full-/Part-Time indicator) | 1 | 1.240877754 | 0.427616255 | 1337929 |
| pubpriv (Public/Private Sector Indicator) | 2 | 1.779814257 | 0.589809704 | 1337929 |
| occ_move (Occupational Move Indicator) | 0 | 0.175693495 | 0.380559182 | 1337929 |
| ind_move (Industry Move Indicator) | 0 | 0.082798963 | 0.27557811 | 1170290 |
| reg_move (Region Move Indicator) | 0 | 0.043395524 | 0.203745803 | 1337929 |

The dataset comprises over 1.3 million person-year observations, drawn from the ASHE longitudinal panel spanning 2011-2023. Due to the panel structure of the dataset, some individuals appear multiple times, but survey weights are used to adjust and ensure the statistics are representative.

The dependent variable, hourly earnings growth (heg), is highly skewed, with a median of 3.02% and a mean of 8.12%, alongside a notably large standard deviation. This confirms considerable variation in wage growth and supports the methodological choice to log-transform the variable in regression analyses, in line with prior literature.

The job switch indicator (sjd) shows that approximately 9% of observations reflect a job change between consecutive years (mean=1.09, where 1=stayer, 2=switcher), highlighting that the sample predominantly comprises of job stayers. This aligns with broader labour market patterns of moderate mobility in the UK, compared to other developed countries (ONS, 2024).

The gender variable (sex) has a mean of 1.49, indicating a slightly male-skewed sample (where male=1, female=2), though the balance is relatively even. The median age is 43, with a fairly normal distribution (mean=42.65;sd=12.55), suggesting that the dataset captures a broad range of working-age individuals.

Other variables provide further insight into employment characteristics. The majority of individuals are employed full-time (ft) (mean=1.24, where 1=full-time, 2=part-time), and the sectoral composition (pubpriv) shows a slight skew towards private sector employment (mean=1.78, where 1=public, 2=private).

Mobility variables reveal that most workers remain in the same occupation (mean=0.18, where 0=same occupation, 1=new occupation), industry (mean=0.08, where 0=same industry, 1=new industry), and region (mean=0.04, where 0=same region, 1=new region) year-to-year. Observations are lower for industry mobility due to the removal of certain industries from the sample to comply with statistical disclosure control requirements.

Together, these summary statistics direct the regression approach, including the use of fixed effects, log transformations, and interaction terms to disentangle the relationship between job mobility, gender and earnings growth.

Table 2 presents the descriptive statistics for the variables used in the analysis, split by gender. This breakdown shows some early patterns that will be explored later in the discussion. Women make up a slightly higher proportion of job switchers than men, but a lower proportion of full-time workers and private sector employees. Gender differences are also evident across occupations and industries, with women being more concentrated in Caring, Leisure and Other Service Occupations and Human, Health and Social Work Activities. Women are also more likely to switch to a new occupation when changing jobs, but less likely to move to a new region. Although these figures are largely descriptive, they highlight the importance of controlling for these factors in the analysis, given that they may affect the relationship between job switching and earnings across genders.

*Table 2: Gender split in categories used in analysis (e.g. 48.8% of Stayers are Female and 51.2% are Male)*

| Category | Female | Male |
|---|---|---|
| **Stayers (%)** | 48.76361 | 51.23639 |
| **Switchers (%)** | 52.05487 | 47.94513 |
| **Full time (%)** | 39.90184 | 60.09816 |
| **Part time (%)** | 77.89151 | 22.10849 |
| **Private (%)** | 40.68299 | 59.31701 |
| **Public (%)** | 65.79224 | 34.20776 |
| **Managers, Directors and Senior Officials (%)** | 35.20661 | 64.79339 |
| **Professional Occupations (%)** | 50.32099 | 49.67901 |
| **Associate Professional Occupations (%)** | 44.04624 | 55.95376 |
| **Administrative and Secretarial Occupations (%)** | 75.96169 | 24.03831 |
| **Skilled Trades Occupations (%)** | 10.8579 | 89.1421 |
| **Caring, Leisure and Other Service Occupations (%)** | 80.86389 | 19.13611 |
| **Sales and Customer Service Occupations (%)** | 63.87087 | 36.12913 |
| **Process, Plant and Machine Operatives (%)** | 12.05681 | 87.94319 |
| **Elementary Occupations (%)** | 45.79358 | 54.20642 |
| **Agriculture, Forestry and Fishing Industry (%)** | 32.27416 | 67.72584 |
| **Manufacturing Industry (%)** | 21.81011 | 78.18989 |
| **Construction Industry (%)** | 18.56798 | 81.43202 |
| **Wholesale and Retail Trade; Repair of Motor Vehicles and Motorcycles (%)** | 45.34782 | 54.65218 |
| **Transportation and Storage Industry (%)** | 21.2577 | 78.7423 |
| **Accommodation and Food Service Activities (%)** | 53.05009 | 46.94991 |
| **Information and Communication Industry (%)** | 29.59532 | 70.40468 |
| **Financial and Insurance Activities (%)** | 46.5386 | 53.4614 |
| **Real Estate Activities (%)** | 52.87226 | 47.12774 |
| **Professional, Scientific and Technical Activities (%)** | 46.10888 | 53.89112 |
| **Administrative and Support Service Activities (%)** | 43.27599 | 56.72401 |
| **Public Administration and Defence; Compulsory Social Security (%)** | 49.19178 | 50.80822 |
| **Education Industry (%)** | 67.44723 | 32.55277 |
| **Human, Health and Social Work Activities (%)** | 77.73246 | 22.26754 |
| **Arts, Entertainment and Recreation Industry (%)** | 48.18905 | 51.81095 |
| **New Occupation (%)** | 51.7744 | 48.2256 |
| **Same Occupation (%)** | 48.4726 | 51.5274 |
| **New Region (%)** | 43.53386 | 56.46614 |
| **Same Region (%)** | 49.30307 | 50.69693 |

### 3.3: Main Regression Specification

The main model is a fixed effects regression, as used by many previous studies (Fuller, 2008; Avram, et al., 2023; Pearlman, 2018), to account for unobserved traits that may affect gender differences in the impact of switching jobs on earnings. Fixed effects control for heterogeneity, allowing a cleaner estimation of the effect of job switching. The model includes an interaction term between job switching and gender, as has been used in previous studies (Avram, et al., 2023), to allow for direct comparisons to be made between the two groups. This approach reduces the possibility of omitted variable bias and addresses potential endogeneity by isolating the effects of job switching on earnings growth.

The model was estimated in R, and is specified as follows:

$$\ln(heg_{it}) = \beta_1 (JobSwitch_{it} \times Sex_{it}) + X'_{it}\beta + \alpha_i + \gamma_t + \varepsilon_{it}$$

In this specification, $\ln(heg_{it})$ represents the log of hourly earnings growth for individual $i$ between $t-1$ and $t$. $JobSwitch_{it} \times Sex_{it}$ is an interaction term capturing the effect of job switching on earnings by gender. $X'_{it}$ represents control variables including age-group, part-time status, sector (public vs private) and occupation. $\alpha_i$ denotes individual fixed effects, and $\gamma_t$ captures year-specific fixed effects to absorb macroeconomic shocks and trends.

The error term $\varepsilon_{it}$ is assumed to be idiosyncratic and heteroskedastic, with standard errors clustered at the individual level to account for repeated observations over time. Survey weights were applied to ensure representativeness. The log transformation of the dependent variable allows coefficients to be interpreted approximately as percentage changes in wage growth.

## 3.4: Factor-Specific Fixed Effects Model

To investigate whether the difference in impact of job switching for men and women varies across demographic and occupational subgroups, a series of factor-specific fixed effects models were estimated. These models interacted the job switch x sex term with additional factors (age-group, part-time status, sector (public vs private), occupation, and occupation change status), to assess whether the gendered wage effects of job mobility differ across these characteristics, as suggested by earlier studies (Pearlman, 2018; Brett & Stroh, 1997; Avram, et al., 2023).

Before estimation, the dataset was filtered to remove observations with missing values in the relevant variables, ensuring consistency. The following fixed effects regressions were run:

$$\ln(heg_{it}) = \beta_1 (JobSwitch_{it} \times Sex_{it} \times AgeGroup_{it}) + \alpha_i + \gamma_t + \varepsilon_{it} \quad (1)$$

$$\ln(heg_{it}) = \beta_1 (JobSwitch_{it} \times Sex_{it} \times FPT_{it}) + \alpha_i + \gamma_t + \varepsilon_{it} \quad (2)$$

$$\ln(heg_{it}) = \beta_1 (JobSwitch_{it} \times Sex_{it} \times Sector_{it}) + \alpha_i + \gamma_t + \varepsilon_{it} \quad (3)$$

$$\ln(heg_{it}) = \beta_1 (JobSwitch_{it} \times Sex_{it} \times Occupation_{it}) + \alpha_i + \gamma_t + \varepsilon_{it} \quad (4)$$

$$\ln(heg_{it}) = \beta_1 (JobSwitch_{it} \times Sex_{it} \times OccupationMove_{it}) + \alpha_i + \gamma_t + \varepsilon_{it} \quad (5)$$

where $AgeGroup_{it}$ represents age group (6 categories), $FPT_{it}$ is a dummy variable for full/part time status, $Sector_{it}$ is a dummy variable for public/private sector, $Occupation_{it}$ refers to one of 9 occupational groups, and $OccupationMove_{it}$ indicates whether an individual changed occupation.

Similarly to the main fixed effects model, each factor-specific model includes individual and year fixed effects, with standard errors clustered at the individual level. Weights were also applied to correct for non-response bias and ensure representativeness.

These triple interaction terms allow us to explore how the gender-job switch relationship varies across different factors. Additional controls were not included to avoid introducing multicollinearity or overcomplicating the specification and to maintain clarity in isolating the interaction across the specific factor of interest.

## 3.5: Robustness Checks

To assess the stability of the results, several robustness checks were conducted. To provide a benchmark for the fixed effects estimates, a pooled OLS model was run with the same set of variables, but no individual controls. The OLS model, which lacks individual and year fixed effects, is expected to yield significantly different coefficients compared to the fixed effects model. This approach is commonly used in wage dynamics literature (Hsiao, 1986) to improve the accuracy of estimates.

To further test the robustness, the main regression was rerun after excluding certain variables, and multicollinearity was assessed using condition numbers. This process led to the exclusion of some redundant or highly collinear variables, specifically industry and region. The model was also re-estimated excluding most control variables to check the stability of the interaction term.

To assess the stability of standard errors, alternative clustering strategies were explored. Whilst individual-level clustering was selected for the main regression due to the panel structure of the data, clustering was also tested by region to account for local labour market variation and by year to account for economic shocks. This approach aimed to determine whether the results were sensitive to different clustering choices.

Finally, condition numbers were checked for factor-specific models to assess potential multicollinearity. This helped to identify any potential inflation of standard errors due to collinearity between explanatory variables.

## 3.6: Limitations

Despite the strengths of the data and methodology, several limitations should be acknowledged. Firstly, the dataset lacks potentially important variables, such as educational attainment, skill level, parenthood status, or voluntary vs involuntary job changes, which have been found to be significant factors in previous research (Avram, et al., 2023; Cominetti, et al., 2022). The absence of these variables introduces a risk of omitted variable bias. Whilst the use of individual fixed effects and clustering helps to mitigate these concerns, they should still be noted for future research.

Secondly, despite the ASHE dataset being a high-quality source of administrative data, it excludes the self-employed and may underrepresent informal employment, limiting the generalisability of the findings to the wider labour market. Additionally, ASHE relies on employer responses, and although response rates are typically higher than in other surveys (such as the LFS), they remain low, particularly during the Covid-19 period (Forth, et al., 2022). Attrition could introduce selection bias, especially given the longitudinal nature of the data, which drops those not surveyed in consecutive years. On average, attrition is around 25% (Forth, et al., 2022). Although longitudinal weights are used to address this, their effectiveness depends on the nature of non-response.

Another limitation concerns the timeliness of the data, as there is a lag between the ASHE reference period and its publication, which limits the ability to capture the most recent labour market dynamics, particularly in such a rapidly changing environment.

Changes in the methodology over time also affect comparability. In particular, the occupational classification system used in ASHE was updated in 2020, transitioning from SOC10 to SOC20. These two coding systems are not fully compatible, especially at detailed occupational levels. To ensure comparability, the analysis was restricted to major occupational groups that are approximately consistent across coding systems.

Lastly, whilst the fixed effects model helps to control for time-invariant individual characteristics and includes a range of factors, it does not fully address potential endogeneity issues such as reverse causality. More advanced causal inference techniques, such as instrumental variable (IV) methods were not implemented in this analysis. Therefore, the results should be interpreted as indicative of associations rather than definitive causal relationships.

# 4: Results

This section presents the empirical findings on how job switching influences hourly earnings growth, with a focus on gender differences. It begins with descriptive trends, then outlines the core fixed effects results, followed by a detailed breakdown of factor-specific results to highlight key patterns.

## 4.1: Descriptive Trends in Hourly Earnings Growth

**Figure 1:** Line Chart of Gender Differences in the Impact of Switching Jobs on Hourly Earnings Growth from 2011-2023

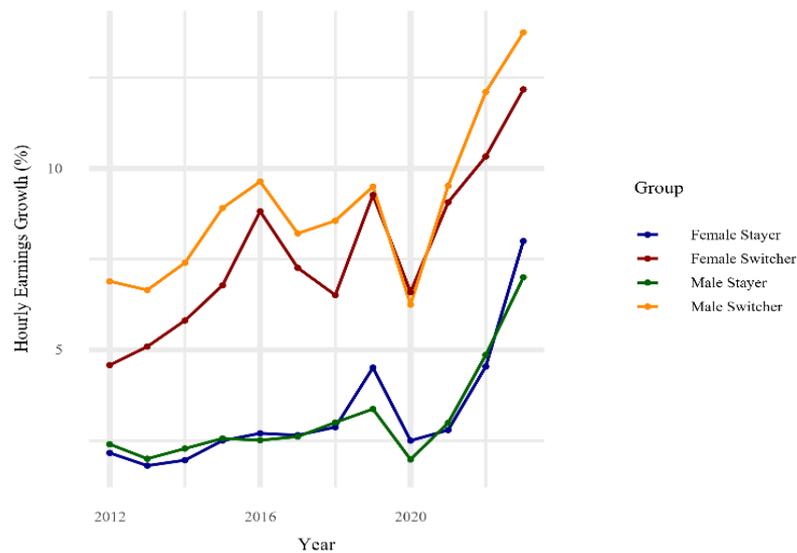

**Figure 2:** Line Chart of Gender Differences in the Impact of Switching Jobs on Hourly Earnings from 2011-2023

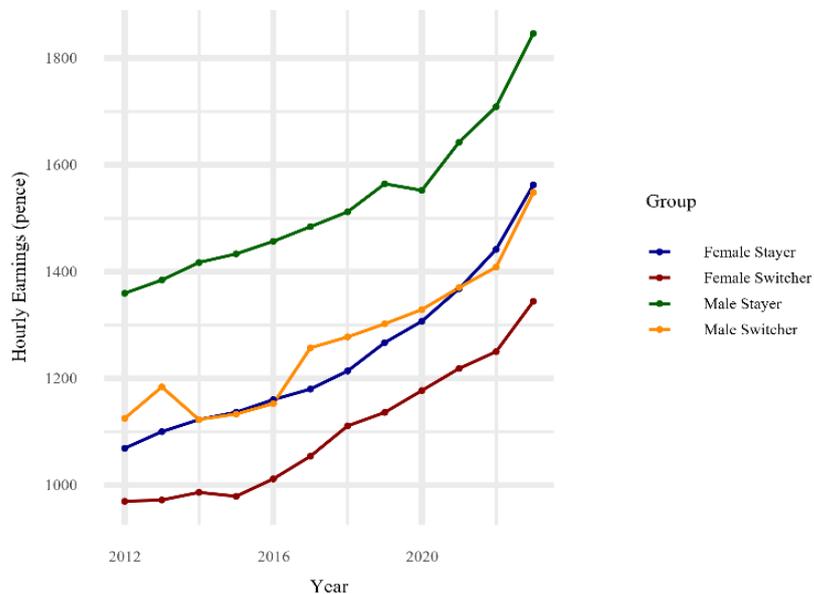

To contextualise regression findings, median hourly earnings and earnings growth were first examined by gender and job-switching status from 2011 to 2023. Figure 1 shows trends in median hourly earnings growth for male and female switchers and stayers (see Appendix C Table 3a for full data). Over this period, job switchers consistently experienced higher earnings growth than stayers, with male switchers typically showing the highest growth rates, except in 2020, when female switchers briefly surpassed them (6.59% vs 6.25%). The gap between stayers and switchers has been more pronounced for men, suggesting stronger wage returns to mobility for male workers. In contrast, earnings growth for stayers remained relatively similar across genders throughout the period.

Although earnings growth is consistently higher for workers who change jobs, job stayers tend to earn a higher hourly wage, as shown in Figure 2 (see Appendix C Table 3b for full data). Male workers also consistently earned more than female workers, but faced a larger gap between stayers and switchers. According to the ONS (2019), this may partly reflect greater skill accumulation and job stability associated with longer tenure. In addition, younger workers, who are more likely to switch jobs, tend to start from a lower wage but experience faster growth.

These patterns highlight the potential impact of job switching on earnings growth and gender disparities. The regression analysis that follows tests these relationships more rigorously, controlling for demographic and job-related factors.

## 4.2: Fixed Effects Regression

This section presents the results of the main fixed effects regression model estimating the gendered impact of job switching on earnings growth, controlling for various factors. The dependent variable is log of hourly earnings growth, so coefficients can be interpreted as approximate percentage changes. Table 3 presents the full regression results, whilst Figure 3 provides a visual summary of the coefficients and 95% confidence intervals.

*Table 3:* Main Regression Results

| Variable | Coefficient | t-value | p-value | [95% conf interval] |
|---|---|---|---|---|
| **Switcher (Binary)** | 0.829039589 *** (0.011602) | 71.45886 | 0 | (0.806301,0.851778) |
| **Male (Binary)** | -0.253025716 (0.160906) | -1.5725 | 0.115835 | (-0.5684,0.062347) |
| **Age 21-24** | -0.097078832*** (0.020207) | -4.80425 | 1.55E-06 | (-0.13668,-0.05747) |
| **Age 25-34** | -0.185957023*** (0.02399) | -7.7513 | 9.13E-15 | (-0.23298,-0.13894) |
| **Age 35-49** | -0.257494148*** (0.029688) | -8.67325 | 4.22E-18 | (-0.31568,-0.19931) |
| **Age 50-64** | -0.220382462*** (0.035182) | -6.2641 | 3.76E-10 | (-0.28934,-0.15143) |
| **Age 65+** | -0.166656036** (0.050792) | -3.28112 | 0.001034 | (-0.26621,-0.0671) |
| **Part-Time (Binary)** | 0.27691779*** (0.010468) | 26.45326 | 5.4E-154 | (0.2564,0.297435) |
| **Public Sector (Binary)** | -0.013871252 (0.017412) | -0.79663 | 0.425667 | (-0.048,0.020257) |
| **Professional Occupations** | 0.054720397* (0.024623) | 2.222335 | 0.026262 | (0.00646,0.102981) |
| **Associate Professional Occupations** | 0.015687317 (0.023474) | 0.668298 | 0.503944 | (-0.03032,0.061695) |
| **Administrative and Secretarial Occupations** | -0.065373884** (0.024025) | -2.72105 | 0.006508 | (-0.11246,-0.01828) |
| **Skilled Trades Occupations** | 0.014708401 (0.03068) | 0.47941 | 0.631648 | (-0.04542,0.074841) |
| **Caring, Leisure and Other Service Occupations** | -0.041230001 (0.026724) | -1.54281 | 0.122879 | (-0.09361,0.011148) |
| **Sales and Customer Service Occupations** | -0.184954734*** (0.02381) | -7.76801 | 8E-15 | (-0.23162,-0.13829) |
| **Process, Plant and Machine Operatives** | -0.021189163 (0.030931) | -0.68504 | 0.49332 | (-0.08181,0.039435) |
| **Elementary Occupations** | -0.172972242*** (0.024438) | -7.07795 | 1.47E-12 | (-0.22087,-0.12507) |
| **Switcher x Male** | 0.096303524*** (0.01736) | 5.547587 | 2.9E-08 | (0.062279,0.130328) |
| **R²** | 0.381771279 | | | |
| **Adjusted R²** | 0.130566993 | | | |
| **N** | 880678 | | | |

Robust standard errors are reported in brackets.
Significance levels were denoted using conventional markers: p<0.001 (***), p<0.01(**), p<0.05(*).

R-squared provides a measure of the proportion of variance in the dependent variables that is explained by the model's independent variables (Miles, 2005). In this model, the R-squared value of 0.382 suggests that about 38.2% of the variation in earnings growth is explained by the model's predictors. When individual and year fixed effects are included to control for unobserved heterogeneity, the adjusted R-squared falls to 0.131. This reflects the trade-off of fixed effects, with reduced explanatory power but greater robustness.

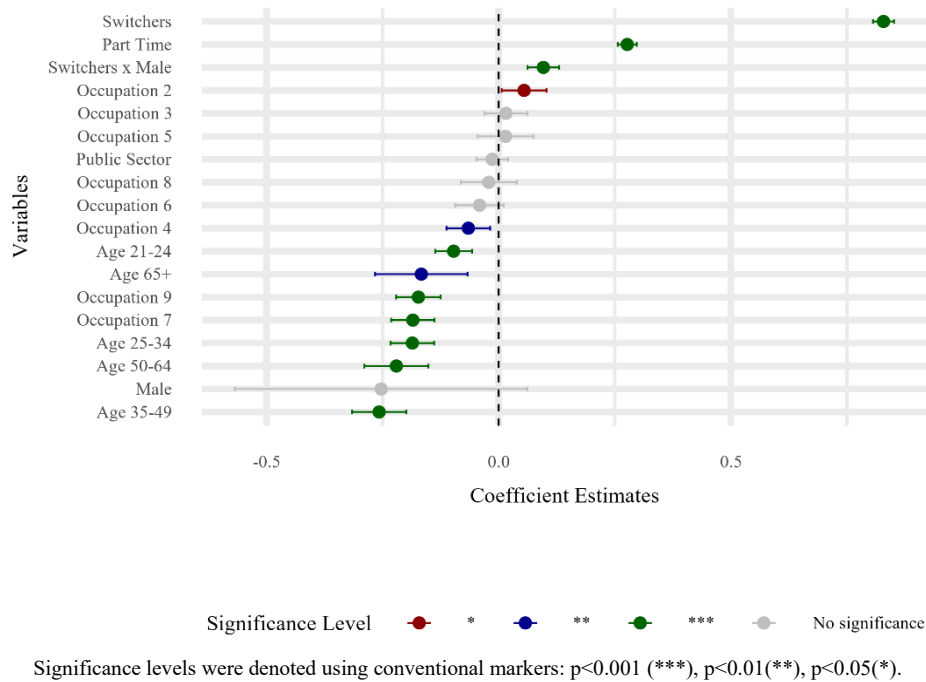

*Figure 3:* Fixed Effects Regression Coefficients Plot

Significance levels were denoted using conventional markers: p<0.001 (***), p<0.01(**), p<0.05(*).

The coefficient for switching jobs is estimated to be 0.829 (p<0.001), indicating a positive and statistically significant relationship with the dependent variable. This suggests that job switchers experience an estimated 82.9% increase in hourly earnings growth compared to non-switchers, holding other factors constant. This finding aligns with previous research which highlights that job mobility is a key mechanism for wage growth, particularly early in the career.

The interaction term between job switching and gender is also positive and statistically significant ($\beta$=0.096, p<0.001), suggesting that male switchers benefit from an additional earnings premium of approximately 9.6%. This implies a gender disparity in the wage returns to job mobility, with men reaping greater rewards from switching jobs than women.

However, being male is associated with a non-significant decrease in hourly earnings growth ($\beta$=-0.253). Although this contrasts with prior studies, this result is not statistically significant, and the large standard error (0.161) implies that the estimate is imprecise and should be interpreted with caution.

Age is a significant predictor of earnings growth, with older workers experiencing lower earnings growth than the reference group (16-20). The largest penalties are seen amongst those aged 35-49 (β=-0.257, p<0.001) and 50-64 (β=-0.220, p<0.001), possibly reflecting longer tenure in roles and reduced job mobility, both associated with lower earnings growth.

Unexpectedly, part time work is associated with a significant positive coefficient (β=0.277, p<0.001), suggesting that part-time workers may experience higher earnings growth compared to full-time workers. This finding differs from most literature, which typically associates part-time work with lower earnings growth. Whilst 77.9% of part-time workers in the sample are female, and prior research often associates women with lower earnings growth, the gender composition alone is unlikely to explain this result. This may be explained by the composition of the sample, but further investigation is needed to better understand the reasons behind this result.

Different occupations show substantial variation. Workers in technical and associate professional roles have significantly higher earnings growth than the reference group (β=0.055, p<0.05), whilst those in sales and customer service occupations experience significant earnings penalties (β=-0.185, p<0.001). These findings are broadly consistent with skill-based segmentation in the labour market. However, many of the results for occupation are not statistically significant, so should be interpreted with caution.

### 4.3: Factor-Specific Fixed Effects Regression

To explore how the gender earnings premium for job switchers varies across demographic and occupational characteristics, additional regressions were estimates with three-way interaction terms between gender, job switching, and selected variables.

*Table 4: Factor-Specific Regression Results*

| Variable | Coefficient | t-value | p-value | [95% conf interval] | Factor |
|---|---|---|---|---|---|
| Switcher x Male | -0.14387* (0.069841) | -2.05993 | 0.039406 | (-0.28075, -0.00698) | Age |
| Switcher x Male x Age 21-24 | 0.069946 (0.083089) | 0.841818 | 0.399891 | (-0.09291, 0.232798) | Age |
| Switcher x Male x Age 25-34 | 0.204632** (0.075322) | 2.716765 | 0.006593 | (0.057003, 0.352262) | Age |
| Switcher x Male x Age 35-49 | 0.321355*** (0.076535) | 4.19877 | 2.68E-05 | (0.171347, 0.471362) | Age |
| Switcher x Male x Age 50-64 | 0.325161*** (0.083577) | 3.890537 | 0.0001 | (0.161351, 0.48897) | Age |
| Switcher x Male x Age 65+ | 0.255907 (0.232733) | 1.099572 | 0.27152 | (-0.20024, 0.712058) | Age |
| Switcher x Male | 0.072033*** (0.019429) | 3.707569 | 0.000209 | (0.033953, 0.110113) | Full/Part Time |
| Switcher x Male x Part-Time | -0.08456 (0.045985) | -1.83893 | 0.065927 | (-0.17469, 0.00556) | Full/Part Time |
| Switcher x Male | 0.07894*** (0.020329) | 3.883079 | 0.000103 | (0.039095, 0.118785) | Public/Private Sector |
| **Switcher x Male x Public Sector** | 0.02063 (0.041458) | 0.497609 | 0.61876 | (-0.06063, 0.101886) | Public/Private Sector |
| **Switcher x Male** | 0.114392 (0.061546) | 1.858638 | 0.06308 | (-0.00624, 0.235022) | Occupation |
| **Switcher x Male x Professional Occupations** | 0.03888 (0.07178) | 0.541659 | 0.588054 | (-0.10181, 0.179566) | Occupation |
| **Switcher x Male x Associate Professional Occupations** | -0.03439 (0.076054) | -0.45216 | 0.651156 | (-0.18345, 0.114676) | Occupation |
| **Switcher x Male x Administrative and Secretarial Occupations** | -0.05683 (0.078333) | -0.72549 | 0.46815 | (-0.21036, 0.0967) | Occupation |
| **Switcher x Male x Skilled Trades Occupations** | -0.15944 (0.139499) | -1.14297 | 0.253053 | (-0.43286, 0.113972) | Occupation |
| **Switcher x Male x Caring, Leisure and Other Service Occupations** | 0.001256 (0.088337) | 0.014214 | 0.988659 | (-0.17188, 0.174393) | Occupation |
| **Switcher x Male x Sales and Customer Service Occupations** | -0.0463 (0.076256) | -0.60716 | 0.543743 | (-0.19576, 0.10316) | Occupation |
| **Switcher x Male x Process, Plant and Machine Operatives** | -0.02898 (0.105349) | -0.27506 | 0.78327 | (-0.23546, 0.177504) | Occupation |
| **Switcher x Male x Elementary Occupations** | 0.041817 (0.074323) | 0.562634 | 0.573685 | (-0.10385, 0.187488) | Occupation |
| **Switcher x Male** | 0.177653*** (0.031074) | 5.717072 | 1.08E-08 | (0.116749, 0.238557) | Occupation Move |
| **Switcher x Male x Same Occupation** | -0.09445* (0.040593) | -2.32687 | 0.019973 | (-0.17402, -0.01489) | Occupation Move |
| **R² (Age)** | 0.381123 | | | | Age |
| **Adjusted R² (Age)** | 0.129648 | | | | Age |
| **R² (Full/Part Time)** | 0.381632 | | | | Full/Part Time |
| **Adjusted R² (Full/Part Time)** | 0.12931 | | | | Full/Part Time |
| **R² (Sector)** | 0.380867 | | | | Public/Private Sector |
| **Adjusted R² (Sector)** | 0.130386 | | | | Public/Private Sector |
| **R² (Occupation)** | 0.381268 | | | | Occupation |
| **Adjusted R² (Occupation)** | 0.129836 | | | | Occupation |
| **R² (Occupation Move)** | 0.382053 | | | | Occupation Move |
| **Adjusted R² (Occupation Move)** | 0.130978 | | | | Occupation Move |
| **N** | 880678 | | | | |

Robust standard errors are reported in brackets.
Significance levels were denoted using conventional markers: p<0.001 (***), p<0.01(**), p<0.05(*).

Male switchers in the 25-34 age group experienced a statistically significant earnings growth premium relative to female switchers in the same group (β=0.205, p<0.01), with the premium increasing further for those aged 35-49 (β=0.321, p<0.001) and 50-64 (β=0.325, p<0.001). This suggests that the gender gap in earnings growth from job switching may widen with age, though effects for the youngest and oldest groups were not statistically significant.

The interaction between male switchers and part-time work was negative and marginally insignificant (β=-0.085, p=0.066), suggesting that the male part-time switchers may receive lower earnings growth than their full-time counterparts, though this result should be interpreted with caution.

The interaction between male switchers and public sector employment was small and not statistically significant (β=0.021, p=0.619), suggesting that the male earnings advantage from switching jobs is relatively consistent across sectors, and that sector of employment is unlikely to be a major factor in shaping gender differences in returns to switching.

When interacting gender and switching status with occupational groups, no statistically significant patterns were observed, and confidence intervals were wide. This indicates limited evidence of variation in the gender premium to switching across occupational groups.

Male job switchers who stay in the same occupation experience smaller earnings growth compared to those who switch to a new occupation. For males who switch jobs within the same occupation, earnings gains compared to females diminish (β=-0.094, p<0.05). This suggests that occupational mobility may be a key driver in returns to switching.

Amongst these models, occupational mobility had the highest explanatory power, with an adjusted R-squared of 0.131. This was followed closely by sector, occupation and age (Adj. $R^2$=0.130 for all), whilst full/part time status contributed slightly less (Adj. $R^2$=0.129). Although these differences are small, they suggest that occupational mobility may be marginally more important in explaining earnings growth. Age also appears to play a meaningful role, consistent with the findings that the gender gap in wage returns to job switching tends to widen with age. Together, these results indicate that both occupational mobility and age may be key contributors to gender differences in the impact of switching jobs on earnings.

### 4.4: Robustness Checks

The fixed effects estimates proved stable across a range of robustness checks (see Appendix B Table 3). Compared to the pooled OLS model, the fixed effects showed notable differences in coefficients, particularly for age and occupation, highlighting the role of unobserved heterogeneity. The explanatory power also improved considerably, with the $R^2$ increasing from 0.049 under OLS to 0.382, underscoring the importance of controlling for individual and time effects when estimating the impact of job switching and gender on earnings growth.

The interaction term between job switching and gender remained positive and statistically significant across alternative specifications where most control variables were excluded. This indicates that the main result is not driven by specific model choices or omitted variable bias.

Standard errors were robust to different clustering strategies. Whilst clustering by region or year increased the standard errors slightly, the key coefficients remained significant. Although industry and region were included in the clustering models, the consistency of results suggests that the main findings are not sensitive to the choice of clustering or model specification, affirming that individual-level clustering, remains the most appropriate choice given the panel structure of the data.

Finally, condition number checks in factor-specific models indicated potential multicollinearity in the age and occupation models, which could inflate standard errors. Consequently, the results from these models should be interpreted with caution. However, the models focused on full/part time status, sector and occupational mobility did not exhibit these issues.

# 5: Discussion

This study investigates gender disparities in the impact of switching jobs on earnings, and whether these vary across occupational and demographic groups, using recent panel data from the Annual Survey of Hours and Earnings (ASHE). The key findings, which are mostly consistent with previous literature, highlight that whilst job switching does generally lead to higher earnings growth, men tend to benefit to a greater extent than women, even after controlling for sector, age, and other factors. These findings shed light on the complex relationship between gender, job switching and earnings.

Although this research faced some methodological limitations due to data constraints, many of the results are statistically significant and provide a robust framework for future research to build upon. The results were largely as expected, though some findings were surprising, offering new insights. These results have important implications for future research and policymaking aimed at reducing gender inequalities in the labour market.

## 5.1: Discussion of Results

Initial analysis of descriptive statistics found that workers who changed jobs consistently experienced higher earnings growth than those who stayed in the same job between 2011 and 2023, with male switchers typically achieving the highest growth rates. The only exception was in 2020, when female switchers briefly outpaced male switchers, likely reflecting pandemic-related shifts in labour market dynamics. These findings are mirrored in the main regression results, which confirmed that job switching is positively associated with earnings growth, a finding consistent with prior research (Cominetti, et al., 2022; Kirkup & Petrie, 2020) that frames switching as an important mechanism for wage growth, particularly for younger workers. However, the presence of a gender difference in the wage returns to switching indicates that the benefits are not evenly distributed, with male switchers experiencing higher earnings growth after controlling for fixed effects. This is in line with prior UK-based studies (ONS, 2022; Avram, et al., 2023), and highlights the ongoing gender inequality in the labour market.

This consistency with existing research underscores the persistence of gender inequalities in the labour market, even amongst job switchers, a group more proactive in wage growth. The descriptive statistics offer additional context. Although women represent a slightly larger share of switchers (52.1%) than stayers (48.8%), they remain disproportionately concentrated in part-time roles (77.9% female) and lower-paid occupations such as Administrative & Secretarial occupations (76%) and Caring, Leisure and Other Service roles (80.9%). Unexpectedly, regression analysis found no statistically significant effects for occupation or part-time status on the gendered returns to switching jobs, suggesting that whilst occupational and working time segregation likely contribute to wider gender inequalities, they do not fully explain the earnings gap between male and female switchers. This partially aligns with findings from Loprest (1992), who similarly concluded that occupational differences do not fully explain gender wage gaps following job changes, but noted that part-time status does partially contribute to the gap. In contrast, other research concluded that occupational segregation plays a bigger role (Rosenfeld & Spenner, 1992). However, it is important to note that both studies were conducted in the US over thirty years ago, so the differences may reflect variations between labour market structures and gender dynamics in the UK and the US. This highlights the importance of contextual factors and suggests that further UK-based research is needed.

Prior research supports this interpretation. Higher likelihood of negotiating wages when switching jobs, better professional networks, and employer biases that favour men (Leibbrandt & List, 2015; Kronberg, 2013; Avram, et al., 2023) are given as potential reasons for gender inequality in the returns to switching jobs in previous literature, and together, these explanations suggest that this inequality is driven less by occupational or working time characteristics and more by how labour market institutions and behaviours respond to gender during the switching process.

Beyond the main effects, the study also examined whether specific factors brought different financial returns. The interaction term between gender, job switching status and sector (public vs private) was positive but not statistically significant, suggesting that sector alone does not substantially shape gender differences, despite evidence that private sector employment is more mobile (ONS, 2019) and male dominated (59.3% male). This finding contradicts expectations that sector would explain part of the gap, and supports the interpretation that biases within switching processes may be relatively consistent across sectors.

On the other hand, age emerged as a significant factor. Male switchers in older age groups experienced progressively larger earnings premiums compared to female switchers, suggesting that gender gaps in returns to switching increases with age. This is consistent with Manning & Swaffield's (2008) findings, and supports cumulative advantage theory (DiPrete & Eirich, 2006), whereby men benefit from early career advantages, accumulating over time and resulting in wider disparities. This study also found that overall wage growth declines with age for both genders, consistent with broader findings on decreasing mobility and wage gains with age (ONS, 2022).

Occupational mobility also played a notable role, as male switchers who changed occupations received higher earnings growth than those who switched jobs in the same occupation, and gender earnings gaps were smaller amongst within-occupation switchers. This is despite marginally more women (51.8%) switching occupations than men (48.8%). One possible explanation for this is that switching occupations could offer access to higher paying roles or industries, especially for individuals with transferable skills or experience that is undervalued in their current occupation. This mostly aligns with the findings of Dex et al. (2008), who found that switching jobs to a lower-status occupation hurt women's earnings more than men's. However, they also concluded that switching to higher-status occupations did not significantly benefit men or women. This discrepancy between this earlier research and the current findings could be explained by the fact that the research is older, and that changes in labour market dynamics, occupational mobility and wage structures over time may have influenced the results. Furthermore, differences in the specific age range of the samples could account for some of the variation in the results.

### 5.3: Practical Implications and Areas for Future Research

The results of this research have important implications for policymakers and employers trying to reduce gender disparities in earnings growth. Previous literature has often focused on the broader scope of gender differences in the impact of switching jobs on earnings, but this study highlights the need for a deeper understanding of the specific factors driving gender earnings disparities in the UK to implement effective policies. Recognising that gender differences in the returns to job switching are not fully explained by occupational and sectoral segregation or working time patterns suggests that interventions should focus on other areas that do influence these gaps, specifically, age and occupational mobility.

Given that the gender gap appears to widen with age, targeted support for women at both early and mid-career stages could be particularly impactful. For example, mentorship programmes aimed at helping early-career women navigate job switching and negotiate salaries could reduce the risk of cumulative disadvantages over time. Mid-to-late career support initiatives such as career transition support and re-skilling opportunities could also help to address widening gaps at later stages of employment.

Additionally, efforts to improve women's access to higher-return occupational moves could include encouraging greater upward occupational mobility through expanded access to training and professional development, alongside specific support for women looking to change occupations. By intervening early and supporting occupational transitions throughout careers, job switching can remain an equal opportunity for women and men.

Although this study offers important insights, it also highlights areas where further research is necessary. Whilst the ASHE dataset provides valuable information on earnings, demographics and occupations, it lacks certain variables that could offer deeper insights. Data on real earnings, ethnicity, and whether job movements were voluntary or involuntary could provide crucial insights into how these factors shape the gender wage gap amongst switchers. Further analysis exploring these factors could help to disentangle the mechanisms that influence earnings growth.

Additionally, future research could benefit from exploring how broader macroeconomic indicators, such as economic cycles, influence gendered returns to job switching. Although this was briefly touched upon in this research, a lack of indicators in the dataset made it hard to draw hard conclusions from this.

Finally, investigating causal relationships between job switching, earnings growth, and the impact of gender on outcomes could benefit future research. By employing more advanced methodologies, such as instrumental variable (IV) techniques, researchers could establish a clear causal link, to provide policymakers with more robust evidence to guide their decision-making, enabling them to design more targeted interventions aimed at reducing gender disparities in mobility related earnings growth.

## 5.5: Conclusion

This study provides new evidence that whilst job switching generally promotes earning growth, the benefits are unevenly distributed by gender. Factors such as age and occupational mobility play an important role in explaining the gender gap in returns to mobility, whilst other structural factors such as occupation, sector and full-/part-time status do not. These findings emphasise the need for targeted policies that address this, to bring more equality to the returns to labour market mobility.

# 6: Conclusion

This dissertation has explored the relationship between gender, job switching and earnings, aiming to identify whether disparities exist and how they are shaped by different factors. The analysis confirmed that men, on average, benefit more from job switching than women, with occupational mobility and age further widening this gap. These findings align with much of the existing literature, but the lack of significant effects from occupation, sector, and working time pattern challenges some previous expectations and highlights areas that warrant further exploration. This unexpected result suggests that traditional explanations of the gap may require re-evaluation within the context of the modern labour market. Understanding these dynamics is crucial for developing targeted interventions aimed at narrowing the gender gap in returns to job switching.

This research contributes to academic debates on gender inequality, labour mobility and earnings by providing updated evidence on how job switching affects earnings differently by gender. It reinforces the idea that mobility is not an equalising force, but instead a process through which inequalities can be reinforced and amplified. The findings may be valuable for policymakers aiming to address the gender pay gap, revealing where disparities are most pronounced and emphasising that promoting job mobility alone is unlikely to eliminate inequalities unless the underlying structures influencing wage returns to mobility are also addressed. In a time of ongoing economic change and evolving work practices, understanding how gender continues to influence job mobility outcomes remains highly relevant. Interventions may need to focus more closely on how opportunities and rewards from mobility are distributed, particularly at different stages of workers' careers. Further research focusing on causal links, such as using instrumental variable approaches, would be useful in deepening our understanding of whether mobility drives earning gaps or simply reflects pre-existing inequalities.

Whilst this study offers new insights, its limitations must be taken into account when considering the findings. The absence of certain variables in the dataset such as education, skill level, and parenthood status, all of which have been highlighted as important factors in existing literature, restricts the depth of the analysis and may obscure important drivers of mobility outcomes. Future research should seek to investigate these factors to develop a more comprehensive understanding of how individual characteristics interact with gender, mobility and earnings. Furthermore, the dataset's exclusion of self-employed and informal workers narrows the generalisability of the results, especially as these forms of work become more prominent in modern economies. Including these groups in future studies would help to build a fuller picture of gender and mobility dynamics.

Additionally, future research would benefit from incorporating real earnings data and macroeconomic indicators to account for the effects of inflation and broader economic shifts, which would help to refine the analysis of gender differences in returns to job switching over time.

Overall, this research lays important groundwork for advancing the understanding of gender differences in the returns to job mobility. Whilst the findings offer valuable insights, they also highlight the need for continued research to better understand the complex and evolving factors that shape gender inequality in the labour market. By developing more comprehensive models that include a broader set of variables, future research can better inform strategies aimed at fostering greater equality in labour market outcomes.

# References

Amior, M. (2019). *Education and Geographical Mobility: The Role of the Job Surplus.* London: Centre for Economic Performance. Retrieved from https://cep.lse.ac.uk/pubs/download/dp1616.pdf

Anderson, T., Forth, J., Metcalf, H., Kirby, S. (2001). *The Gender Pay Gap.* Retrieved from https://www.researchgate.net/profile/Simon-Kirby-2/publication/242534277_Final_Report_to_the_DfEE/links/54ba9c0c0cf253b50e2d040c/Final-Report-to-the-DfEE.pdf

Apergis, N., & Lynch, N. (2022). The impact of economic freedom on the gender pay gap: evidence from a survey of UK households. *Journal of Economic Studies, 49*(1), 61-76. Retrieved from https://www.emerald.com/insight/content/doi/10.1108/jes-09-2020-0444/full/html?casa_token=nP6Kn0F8XMwAAAAA:ZBBYDFsPplizlsY1DXxlrMQSMiIf05KK9H5GwCMtPMb-m9C-vi5YEb-RAlp3A0_vgpqnnPdoxz8ZZkfeYkCus6nif3qSvBkoGo8ujFYp-qcm8uIGM95Y

Avram, S., Harkness, S., Popova, D. (2023). *Gender differences in job mobility and pay progression in the UK.* Institute for Social & Economic Research. Retrieved from https://www.iser.essex.ac.uk/wp-content/uploads/files/working-papers/iser/2023-02.pdf

Babcock, L., & Laschever, S. (2007). *Women Don't Ask: The High Cost of Avoiding Negotiation and Positive Strategies for Change.* New York City: Bantam.

Becker, G. (1964). *Human Capital: A Theoretical and Empirical Analysis, with Special Reference to Education.* New York: Colombia University Press.

Booth, A., & Francesconi, M. (2000). Job mobility in 1990s Britain: Does gender matter? *Research in Labor Economics, 19*, 173-189. Retrieved from https://www.emerald.com/insight/content/doi/10.1016/s0147-9121(00)19008-9/full/html

Bosworth, D., & Kersley, H. (2015). *Opportunities and outcomes in education and work: Gender effects.* UK Commission for Employment and Skills. Retrieved from https://assets.publishing.service.gov.uk/media/5a815349ed915d74e33fd95d/UKCES_Gender_Effects.pdf

Brett, J., & Stroh, L. (1997). Jumping ship: Who benefits from an external labor market career strategy? *Journal of Applied Psychology, 82*(3), 331-341. doi:10.1037/0021-9010.82.3.331

Campbell, K. (1988). Gender Differences in Job-Related Networks. *Work and Occupations, 15*(2), 179-200. Retrieved from https://journals.sagepub.com/doi/10.1177/0730888488015002003

Card, D., Cardoso, A. R., Kline, P. (2015). *BARGAINING, SORTING, AND THE GENDER WAGE GAP: QUANTIFYING THE IMPACT OF FIRMS ON THE RELATIVE PAY OF WOMEN*. Cambridge, MA. Retrieved from https://www.nber.org/system/files/working_papers/w21403/w21403.pdf

Carrillo-Tudela, C., Hobijn, B., She, P., Visschers, L. (2016, May). The extent and cyclicality of career changes: Evidence for the U.K. *European Economic Review, 84*, 18-41. Retrieved from https://www.sciencedirect.com/science/article/pii/S0014292115001464

Cha, Y. (2014, May 2). Job Mobility and the Great Recession: Wage Consequences by Gender and Parenthood. *Sociological Science, 1*(12). Retrieved from https://sociologicalscience.com/articles-vol1-12-159/

CIPD. (2023, May 25). *Flexible and hybrid working practices in 2023.* Retrieved from CIPD: https://www.cipd.org/globalassets/media/knowledge/knowledge-hub/reports/2023-pdfs/2023-flexible-hybrid-working-practices-report-8392.pdf

Cominetti, N., Costa, R., Eyles, A., Moev, T., Ventura, G. (2022). *Changing jobs? Change in the UK labour market and the role of worker mobility.* Resolution Foundation.

Crowley, L. (2024). *The changing face of the youth labour market.* London: Chartered Institute of Personnel and Development.

Dawson, C. (2023). Gender differences in optimism, loss aversion and attitudes towards risk. *British Journal of Psychology, 114*(4), 928-944.

Del Bono, E., & Vuri, D. (2010, July 1). Job mobility and the gender wage gap in Italy. *Labour Economics, 18*(1), 130-142. Retrieved from https://www.sciencedirect.com/science/article/pii/S092753711000076X

Dex, S. W., Joshi, H. (2008, June). Gender differences in occupational wage mobility in the 1958 cohort. *Work Employment and Society, 22*(2), 263-280. Retrieved from https://www.researchgate.net/publication/249828637_Gender_Differences_in_Occupational_Wage_Mobility_in_the_1958_Cohort

Dias, M., Joyce, R., Parodi, F. (2018). *Wage progression and the gender wage gap: the causal impact of hours of work.* The Institute for Fiscal Studies.

DiPrete, T., & Eirich, G. (2006). Cumulative Advantage as a Mechanism for Inequality: A Review of Theoretical and Empirical Developments. *Annual Review of Sociology, 32*, 271-297.

Forth, J., Phan, V., Stokes, L., Bryson, A., Ritchie, F., Whittard, D., & Singleton, C. (2022). *METHODOLOGY PAPER: Longitudinal attrition in ASHE.* Wage & Employment Dynamics.

Francis-Devirle, B., Zaidi, K., Murray, A. (2025). *Women and the UK economy.* House of Commons Library. Retrieved from https://researchbriefings.files.parliament.uk/documents/SN06838/SN06838.pdf

Fredriksson, P., Hensvik, L., Nordström Skans, O. (2018, November). Mismatch of Talent: Evidence on Match Quality, Entry Wages, and Job Mobility. *American Economic Review, 108*(11), 3303-3338. Retrieved from https://docs.iza.org/dp9585.pdf

Fuller, S. (2008, February). Mobility and Wage Trajectories for Men and Women in the United States. *American Sociological Review, 73*(1), 158-183. Retrieved from https://www.jstor.org/stable/pdf/25472518.pdf

Garda, P. (2017, March 30). *Employment ins and outs in OECD countries*. Retrieved from OECD ECOSCOPE: https://oecdecoscope.blog/2017/03/30/employment-ins-and-outs-in-oecd-countries/?utm_source=chatgpt.com

Gottschalk, P. (2001). *Wage Mobility within and between Jobs.* Retrieved from https://www.researchgate.net/publication/4795530_Wage_Mobility_within_and_between_Jobs

Gratton, L., & Cable, D. (2022, June 13). *What's driving the great resignation*. Retrieved from London Business School: https://www.london.edu/think/whats-driving-the-great-resignation

Haltiwanger, J., Hyatt, H., Kahn, L., McEntarfer, E. (2018, April). Cyclical Job Ladders by Firm Size and Firm Wage. *American Economic Journal: Macroeconomics, 10*(2), 52-85. Retrieved from https://www.aeaweb.org/articles?id=10.1257/mac.20150245

Hospido, L. (2009, January). Gender differences in wage growth and job mobility of young workers in Spain. *Investigaciones Económicas, 33*(`), 5-37. Retrieved from https://www.researchgate.net/publication/23954502_Gender_differences_in_wage_growth_and_job_mobility_of_young_workers_in_Spain

Hsiao, C. (1986). *Analysis of Panel Data.* Cambridge: Cambridge University Press.

Jovanovic, B. (1979, October). Job Matching and the Theory of Turnover. *Journal of Political Economy, 87*, 972-990. Retrieved from https://www.jstor.org/stable/pdf/1833078.pdf

Keith, K., & McWilliams, A. (1999). The Returns to Mobility and Job Search by Gender. *ILR Review, 52*(3), 460-477.

Kidd, M. P., O'Leary, N., Sloane, P. (2017, April). The impact of mobility on early career earnings: A quantile regression approach for UK graduates. *Economic Modelling, 62*, 90-1012. Retrieved from https://www.sciencedirect.com/science/article/pii/S0264999317300858?casa_token=Jun-0WYxz3oAAAAA:XFzCdOZ3HPvw2FMky-gvqIJahXu6A3S3TykfNxTC5pSkcUolWB0tv3bmrG0YR_454ETADHpuci0

Kirkup, J., & Petrie, K. (2020). *Job switching and wage growth for low-income workers.* London: Social Market Foundation. Retrieved from https://www.smf.co.uk/wp-content/uploads/2020/11/Job-switching-and-wage-growth-Nov-2020.pdf

Klimczuk, A., & Klimczuk-Kochańska, M. (2016). Dual labor market. In N. Naples, R. C. Hoogland, M. Wickramasinghe, & W. C. Wong, *The Wiley-Blackwell encyclopedia of gender and sexuality studies* (pp. 1-3). Wiley.

Kronberg, A.-K. (2013). Stay or Leave? Externalization of Job Mobility and the Effect on the U.S. Gender Earnings Gap, 1979-2009. *Social Forces, 91*(4), 1117-1146.

Lagerström, J., & Eriksson, S. (2008, November). The Labor Market Consequences of Gender Differences in Job Search. *Journal of Labor Research, 33*. doi:10.1007/s12122-012-9132-2

Le Grand, C., & Tåhlin, M. (2002, December 1). Job Mobility and Earnings Growth. *European Sociological Review, 18*(4), 381-400. Retrieved from https://academic.oup.com/esr/article/18/4/381/551560?login=true

Leibbrandt, A., & List, J. A. (2015, September). Do Women Avoid Salary Negotiations? Evidence from a Large-Scale Natural Field Experiment. *Management Science, 61*(9), 2016-2024. Retrieved from https://www.jstor.org/stable/24551582?seq=2

Loprest, P. (1992). Gender Differences in Wage Growth and Job Mobility. *American Economic Review, 82*(2), 526-532. Retrieved from http://kumlai.free.fr/RESEARCH/THESE/TEXTE/MOBILITY/segmented/Gender%20different%20in%20wage%20growth%20and%20job%20mobility.pdf

Lundberg, S., & Rose, E. (2000, November). Parenthood and the earnings of married men and women. *Labour Economics, 7*(6), 689-710. Retrieved from https://www.sciencedirect.com/science/article/pii/S0927537100000208

Manning, A. (2003). *Monopsony in Motion.* Princeton University Press.

Manning, A., & Robinson, H. (2004, April). Something in the Way She Moves: A Fresh Look at an Old Gap. *Oxford Economic Papers, 56*(2), 169-188. Retrieved from https://www.jstor.org/stable/3488821

Manning, A., & Swaffield, J. (2008, July 1). The Gender Gap in Early-Career Wage Growth. *The Economic Journal, 118*(530), 983-1024. Retrieved from https://academic.oup.com/ej/article/118/530/983/5088806?login=true

Marks, M. A., & Harold, C. (2011, April). Who asks and who receives in salary negotiation. *Journal of Organizational Behavior, 32*(3), 371-394. Retrieved from https://www.researchgate.net/publication/230048548_Who_asks_and_who_receives_in_salary_negotiation

McGee, A., McGee, P., Pan, J. (2015, March). Performance pay, competitiveness, and the gender wage gap: Evidence from the United States. *Economics Letters, 128*, 35-38.

Retrieved from
https://www.sciencedirect.com/science/article/pii/S0165176515000142

McKinsey. (2022, July 13). *The Great Attrition is making hiring harder. Are you searching the right talent pools?* Retrieved from McKinsey&Company: https://www.mckinsey.com/capabilities/people-and-organizational-performance/our-insights/the-great-attrition-is-making-hiring-harder-are-you-searching-the-right-talent-pools

Miles, J. (2005, October 15). *Encyclopedia of Statistics in Behavioral Science.* Retrieved from Wiley Online Library: https://onlinelibrary.wiley.com/doi/abs/10.1002/0470013192.bsa526

Murphy, K., & Welch, F. (1990, April). Empirical Age-Earnings Profiles. *Journal of Labor Economics, 8*(2), 202-229. Retrieved from https://www.jstor.org/stable/2535096?seq=1

Oaxaca, R. (1973, October). Male-Female Wage Differentials in Urban Labor Markets. *International Economic Review, 14*(3), 693-709. Retrieved from https://www.jstor.org/stable/2525981?seq=1

OECD. (2024). *Promoting Better Career Mobility for Longer Working Lives in the United Kingdom.* OECD. Retrieved from https://www.oecd.org/en/publications/promoting-better-career-mobility-for-longer-working-lives-in-the-united-kingdom_2b41ab8e-en/full-report/career-mobility-in-the-united-kingdom-evidence-and-insights-from-recent-trends_52fc8c23.html

ONS. (2017, October 26). *Guide to interpreting Annual Survey of Hours and Earnings (ASHE) estimates*. Retrieved from ONS: https://www.ons.gov.uk/employmentandlabourmarket/peopleinwork/earningsandworkinghours/methodologies/guidetointerpretingannualsurveyofhoursandearningsasheestimates

ONS. (2019, April 29). *Analysis of job changers and stayers*. Retrieved from ONS: http://ons.gov.uk/economy/nationalaccounts/uksectoraccounts/compendium/economicreview/april2019/analysisofjobchangersandstayers

ONS. (2022, April 27). *Comparison of labour market data sources*. Retrieved from ONS: https://www.ons.gov.uk/employmentandlabourmarket/peopleinwork/employmentandemployeetypes/methodologies/comparisonoflabourmarketdatasources

ONS. (2022, May 19). *Job changers and stayers, understanding earnings, UK: April 2012 to April 2021*. Retrieved from ONS: https://www.ons.gov.uk/employmentandlabourmarket/peopleinwork/earningsandworkinghours/articles/jobchangersandstayersunderstandingearningsukapril2012toapril2021/april2012toapril2021

ONS. (2024, November 12). *Dataset: X02: Labour Force Survey flows estimates.* Retrieved from Office for National Statistics: https://www.ons.gov.uk/employmentandlabourmarket/peopleinwork/employmentandemployeetypes/datasets/labourforcesurveyflowsestimatesx02/current

ONS. (2024, October 29). *Gender pay gap in the UK: 2024.* Retrieved from ONS: https://www.ons.gov.uk/employmentandlabourmarket/peopleinwork/earningsandworkinghours/bulletins/genderpaygapintheuk/2024#:~:text=In%20April%202024%2C%20the%20gender%20pay%20gap%20for%20full%2Dtime,more%20than%20double%20at%209.1%25.

ONS. (2025, March 20). *A05 SA: Employment, unemployment and economic inactivity by age group (seasonally adjusted).* Retrieved from ONS: https://www.ons.gov.uk/employmentandlabourmarket/peopleinwork/employmentandemployeetypes/datasets/employmentunemploymentandeconomicinactivitybyagegroupseasonallyadjusteda05sa/current

ONS. (2025). Annual Survey of Hours and Earnings, 1997-2024: Secure Access. *26th Edition*. UK Data Service. doi:http://doi.org/10.5255/UKDA-SN-6689-25

ONS. (2025, February 18). *EMP13: Employment by industry.* Retrieved from ONS: https://www.ons.gov.uk/employmentandlabourmarket/peopleinwork/employmentandemployeetypes/datasets/employmentbyindustryemp13

O'Shea, P. G., & Bush, D. F. (2002, January). Negotiation for Starting Salary: Antecedents and Outcomes Among Recent College Graduates. *Journal of Business and Psychology, 16*(3), 365-382. Retrieved from https://www.researchgate.net/publication/226439084_Negotiation_for_Starting_Salary_Antecedents_and_Outcomes_Among_Recent_College_Graduates

Pearlman, J. (2018). Gender differences in the impact of job mobility on earnings: The role of occupational segregation. *Social Science Research, 74*, 30-44. Retrieved from https://www.sciencedirect.com/science/article/abs/pii/S0049089X17304660?via%3Dihub

Perales, F. (2010). *Occupational Feminization, Specialized Human Capital and Wages: Evidence from the British Labour Market.* University of Essex. Institute for Social and Economic Research. Retrieved from https://www.iser.essex.ac.uk/wp-content/uploads/files/working-papers/iser/2010-31.pdf

Pilossoph, L., Ryngaert, J. M., Wedewer, J. (2024, September). *The Search Costs of Inflation.* Duke University; University of Notre Dame. European Central Bank. Retrieved from https://www.ecb.europa.eu/press/conferences/shared/pdf/20240919_ARC/Ryngaert_paper.pdf

Quintana-Garcia, C., & Elvira, M. M. (2017). The Effect of the External Labor Market on the Gender Pay Gap among Executives. *ILR Review, 70*(1), 132-159. doi:https://doi.org/10.1177/0019793916668529

Ridgeway, C., & Correll, S. (2004, November 8). Motherhood as a Status Characteristic. *Journal of Social Issues, 60*(4), 683-700. Retrieved from https://spssi.onlinelibrary.wiley.com/doi/abs/10.1111/j.0022-4537.2004.00380.x

Rosen, S. (1986). The theory of equalizing differences. In S. Rosen, *Handbook of Labor Economics* (Vol. 1, pp. 641-692).

Rosenfeld, R., & Spenner, K. (1992). Occupational Sex Segregation and Women's Early Career Job Shifts. *Work and Occupations, 19*(4), 424-449. Retrieved from https://journals.sagepub.com/doi/abs/10.1177/0730888492019004005

Rubery, J., Bi-Swinglehurst, I., Rafferty, A. (2024). *Part-time work and productivity.* The Productivity Institute.

Strathmann, L. (2025, February 28). *When People Switch Jobs, What Does It Really Mean for the Economy?* Retrieved from Yale University: https://economics.yale.edu/news/250228/when-people-switch-jobs-what-does-it-really-mean-economy

Topel, R. H., & Ward, M. P. (1992, May). Job Mobility and the Careers of Young Men. *The Quarterly Journal of Economics, 107*(2), 439-479. Retrieved from https://www.jstor.org/stable/2118478?seq=35

Van Ours, J. (1990, December). An International Comparative Study on Job Mobility. *Labour, 4*(3), 33-56. Retrieved from https://onlinelibrary.wiley.com/doi/abs/10.1111/j.1467-9914.1990.tb00020.x

Willetts, D., Alakeson, V., Barker, K., Bell, T., Faribairn, C., Filkin, G., . . . Wilson, N. (2018). *A NEW GENERATIONAL CONTRACT The final report of the Intergenerational Commission.* Resolution Foundation. Retrieved from https://www.resolutionfoundation.org/advanced/a-new-generational-contract/

Women and Equalities Committee. (2021). *Unequal impact? Coronavirus and the gendered economic impact.* House of Commons.

Yankow, J., & Horney, M.-J. (2013, January). Gender DIfferences in Employed Job Seach: Why Do Women Search Less than Men? *Modern Economy, 4*(7), 489-500. Retrieved from https://www.researchgate.net/publication/273743875_Gender_Differences_in_Employed_Job_Search_Why_Do_Women_Search_Less_than_Men

# Appendix

**Appendix A- Variable Reference Table**

**Table 1a.** Reference table for occupation

| Code | Occupation |
|---|---|
| **Occupation 1** | Managers, Directors and Senior Officials |
| **Occupation 2** | Professional Occupations |
| **Occupation 3** | Associate Professional Occupations |
| **Occupation 4** | Administrative and Secretarial Occupations |
| **Occupation 5** | Skilled Trades Occupations |
| **Occupation 6** | Caring, Leisure and Other Service Occupations |
| **Occupation 7** | Sales and Customer Service Occupations |
| **Occupation 8** | Process, Plant and Machine Operatives |
| **Occupation 9** | Elementary Occupations |

**Table 1b.** Reference table for industry

| Code | Occupation |
|---|---|
| **Industry C** | Manufacturing |
| **Industry F** | Construction |
| **Industry G** | Wholesale and retail trade; repair of motor vehicles and motor cycles |
| **Industry H** | Transportation and storage |
| **Industry I** | Accommodation and food service activities |
| **Industry J** | Information and communication |
| **Industry K** | Financial and insurance activities |
| **Industry L** | Real estate activities |
| **Industry M** | Professional, scientific and technical activities |
| **Industry N** | Administrative and support service activities |
| **Industry O** | Public administration and defence; compulsory social security |
| **Industry P** | Education |
| **Industry Q** | Human health and social work activities |
| **Industry R** | Arts, entertainment and recreation |

## Appendix B- Robustness Tests

**Table 2.** Robustness tests comparing the main regression (FE (Main)) with ones with reduced control variables (FE (Reduced Controls)), ones with different clusters (FE (Cluster by Year) and FE (Cluster by Region)) and OLS.

| | FE (Main) | FE (Reduced Controls) | FE (Cluster by Year) | FE (Cluster by Region) | OLS |
|---|---|---|---|---|---|
| **Dependent Variable** | log(heg) | log(heg) | log(heg) | log(heg) | log(heg) |
| **Switcher (Binary)** | 0.8290*** (0.0116) | 0.8557*** (0.0108) | 0.8184*** (0.0368) | 0.8184*** (0.0335) | 0.8935*** (0.0108) |
| **Male (Binary)** | -0.2530 (0.1609) | -0.2200 (0.1445) | -0.2855. (0.1519) | -0.2855. (0.1317) | -0.0239*** (0.0051) |
| **Age 21-24** | -0.0971*** (0.0202) | | -0.0757 (0.0458) | -0.0757** (0.0199) | -0.2117*** (0.0155) |
| **Age 25-34** | -0.1860*** (0.0240) | | -0.1365 (0.0772) | -0.1365** (0.0349) | -0.4448*** (0.0138) |
| **Age 35-49** | -0.2575*** (0.0297) | | -0.1937. (0.0861) | -0.1937*** (0.0381) | -0.8311*** (0.0136) |
| **Age 50-64** | -0.2204*** (0.0352) | | -0.1424. (0.0742) | -0.1424** (0.0383) | -1.032*** (0.0137) |
| **Age 65+** | -0.1667** (0.0508) | | -0.0825 (0.0784) | -0.0825 (0.0552) | -1.105*** (0.0188) |
| **Part-Time (Binary)** | 0.2769*** (0.0105) | | 0.2778*** (0.0241) | 0.2778*** (0.0154) | 0.1308*** (0.0057) |
| **Public Sector (Binary)** | -0.0139 (0.0174) | | 0.0255 (0.0799) | 0.0255 (0.0422) | -0.1546*** (0.0052) |
| **Professional Occupations** | 0.0547* (0.0246) | | 0.0846** (0.0257) | 0.0846* (0.0318) | 0.1077*** (0.0082) |
| **Associate Professional Occupations** | 0.0157 (0.0235) | | 0.0232 (0.0398) | 0.0232 (0.0296) | -0.0135 (0.0086) |
| **Administrative and Secretarial Occupations** | -0.0654** (0.0240) | | -0.0455 (0.0272) | -0.0455 (0.0254) | -0.1383*** (0.0092) |
| **Skilled Trades Occupations** | 0.0147 (0.0307) | | 0.0304 (0.0596) | 0.0304 (0.0382) | 0.1510*** (0.0104) |
| **Caring, Leisure and Other Service Occupations** | -0.0412 (0.0267) | | -0.0534 (0.0303) | -0.0534. (0.0255) | 0.2052*** (0.0106) |
| **Sales and Customer Service Occupations** | -0.1850*** (0.0238) | | -0.00627224 | -0.1304*** (0.0266) | -0.0030 (0.0104) |
| **Process, Plant and Machine Operatives** | -0.0212 (0.0309) | | -0.0063 (0.0517) | -0.0063 (0.0363) | 0.2266*** (0.0110) |
| **Elementary Occupations** | -0.1730*** (0.0244) | | -0.1327*** (0.0245) | -0.1327** (0.0295) | 0.1236*** (0.0098) |
| **Switcher x Male** | 0.0963*** (0.0174) | 0.1029*** (0.0165) | 0.0811** (0.0217) | 0.0811* (0.0280) | 0.0859*** (0.0153) |
| **Manufacturing** | | | 0.0008 (0.0602) | 0.0008 (0.1390) | |
| **Construction** | | | 0.2292** (0.0507) | 0.2292 (0.1477) | |
| **Wholesale and retail trade; repair of motor vehicles and motor cycles** | | | -0.00780208 | -0.1364 (0.1216) | |
| **Transportation and storage** | | | 0.0300 (0.0768) | 0.0300 (0.1368) | |
| **Accommodation and food service activities** | | | -0.1767** (0.0490) | -0.1767 (0.1258) | |
| **Information and communication** | | | -0.0703 (0.1112) | -0.0703 (0.1399) | |
| **Financial and insurance activities** | | | -0.0264 (0.0849) | -0.0264 (0.1298) | |
| **Real estate activities** | | | -0.1150 (0.1739) | -0.1150 (0.1484) | |
| **Professional, scientific and technical activities** | | | 0.0561 (0.0710) | 0.0561 (0.1284) | |
| **Administrative and support service activities** | | | 0.0743 (0.0508) | 0.0743 (0.1325) | |
| **Public administration and defence; compulsory social security** | | | 0.0596 (0.0928) | 0.0596 (0.1593) | |
| **Education** | | | -0.0027 (0.0867) | -0.0027 (0.1414) | |
| **Human health and social work activities** | | | 0.0607 (0.0683) | 0.0607 (0.1346) | |
| **Arts, entertainment and recreation** | | | -0.1541. (0.0730) | -0.1541 (0.1541) | |
| **Region (numeric)** | | | 0.0106 (0.0072) | 0.0106. (0.0053) | |
| **Constant** | | | | | 2.297*** |
| | | | | | (0.0153) |
| **Fixed-Effects:** | --------------- | --------------- | --------------- | --------------- | --------------- |
| **Person identifier (piden)** | Yes | Yes | Yes | Yes | No |
| **Year** | Yes | Yes | Yes | Yes | No |
| _____________________ | _________ | _______ | _______ | _____ | ____ |
| **S.E. type** | by: piden | by: piden | by: year | by: Region | IID |
| **Observations** | 880,678 | 962,705 | 735,713 | 735,713 | 880,678 |
| **R2** | 0.38177 | 0.37408 | 0.41983 | 0.41983 | 0.0486 |
| **Within R2** | 0.0146 | 0.01369 | 0.01806 | 0.01806 | -- |

Robust standard errors are reported in brackets.
Significance levels were denoted using conventional markers: p<0.001 (***), p<0.01(**), p<0.05(*).

**Appendix C- Median Hourly Earnings (Growth) Data Tables**

**Table 3a.** Underlying data for Figure 1 (Line Chart of Gender Differences in the Impact of Switching Jobs on Hourly Earnings Growth from 2011-2023). Table showing median hourly earnings growth for male and female job stayers and switchers, from 2012 to 2023.

| Year | Female Stayers | Female Switchers | Male Stayers | Male Switchers |
|---|---|---|---|---|
| 2012 | 2.16 | 4.58 | 2.4 | 6.89 |
| 2013 | 1.81 | 5.09 | 2 | 6.65 |
| 2014 | 1.96 | 5.81 | 2.28 | 7.4 |
| 2015 | 2.5 | 6.78 | 2.56 | 8.91 |
| 2016 | 2.7 | 8.82 | 2.51 | 9.64 |
| 2017 | 2.65 | 7.26 | 2.61 | 8.21 |
| 2018 | 2.87 | 6.51 | 3 | 8.56 |
| 2019 | 4.51 | 9.27 | 3.37 | 9.5 |
| 2020 | 2.5 | 6.59 | 1.98 | 6.25 |
| 2021 | 2.79 | 9.07 | 2.99 | 9.52 |
| 2022 | 4.54 | 10.33 | 4.86 | 12.11 |
| 2023 | 8 | 12.18 | 7 | 13.75 |

**Table 3b.** Underlying data for Figure 2 (Line Chart of Gender Differences in the Impact of Switching Jobs on Hourly Earnings from 2011-2023). Table showing median hourly earnings for male and female job stayers and switchers, from 2012 to 2023.

| Year | Female Stayers | Female Switchers | Male Stayers | Male Switchers |
|---|---|---|---|---|
| 2012 | 1068.93 | 969.48 | 1359.39 | 1125 |
| 2013 | 1100.06 | 971.98 | 1384.16 | 1183.84 |
| 2014 | 1122.63 | 986.19 | 1417.08 | 1122.7 |
| 2015 | 1136.41 | 978.9 | 1433.21 | 1133 |
| 2016 | 1160 | 1011.56 | 1456.66 | 1152.59 |
| 2017 | 1179.82 | 1054 | 1484.13 | 1257.1 |
| 2018 | 1213.69 | 1110.78 | 1511.89 | 1277.46 |
| 2019 | 1266.85 | 1136 | 1564.44 | 1302.08 |
| **2020** | 1306.93 | 1177.07 | 1552.24 | 1328.84 |
| **2021** | 1367.71 | 1218.32 | 1642.38 | 1369.96 |
| **2022** | 1441.73 | 1250 | 1709 | 1408.34 |
| **2023** | 1562.5 | 1344 | 1846.3 | 1548.1 |